\documentclass[11pt,a4paper,showpacs,english,superscriptaddress,nofootinbib]{revtex4}
\pdfoutput=1
\usepackage{color}
\usepackage{amssymb}
\usepackage{amsmath}
\usepackage{graphicx}
\usepackage{babel}
\usepackage[T1]{fontenc}
\usepackage{hyperref}

\usepackage{amsfonts}

\begin{document}
\title{Exploring Elko typical signature}

\author{M. Dias}
\email{mafd@cern.ch}
\affiliation{Departamento de Ci\^encias Exatas e da Terra,Universidade Federal de S\~ao Paulo\\
Diadema-SP-Brazil}

\author{F. de Campos}
\email{camposc@feg.unesp.br}

\author{J. M. Hoff da Silva} 
\email{hoff@feg.unesp.br; hoff@ift.unesp.br}
\affiliation{Departamento de F\'isica e Qu\'imica, Universidade Estadual Paulista,\\
 Guaratinguet\'a-SP-Brazil}


\begin{abstract}
We study the prospects of observing the presence of a relatively light Elko particle as a possible dark matter candidate,  by pointing out a typical signature for the process encompassing the 
Elko non-locality, exploring some consequences of the unusual Elko propagator behavior when analyzed outside the Elko axis of propagation. 
We also consider the production of a light Elko associated to missing energy and isolated leptons at the LHC, with center of mass energy of 7 and 14 TeV and total luminosity from $1 fb^{-1}$ to $10 fb^{-1}$. 
Basically, the Elko non locality engenders a peculiar signal in the missing energy turning it sensible to the angle of detection.  
\end{abstract}

\maketitle
\noindent

\section{Introduction}

Elko spinor fields are unexpected spin one-half matter fields endowed with mass dimension $1$ \cite{elko,elko2}. Since its recent theoretical discovery, it has attracted much attention, in part by the wide range of possibility opened by such peculiar matter fields in cosmology and physics \cite{COSMO} and in part from the mathematical point of view \cite{MATE}. The word Elko is the acronym for {\it Eigenspinoren des Ladungskonjugationsoperators} or Dual-helicity eigenspinors of the charge conjugation operator (see Eq. (\ref{e2})). 

The two aforementioned characteristics of Elko (namely, spin one-half and mass dimension $1$) makes quite reduced the possible coupling to the Standard Model fields. In fact, keeping in mind that interaction terms with mass dimension greater than four should be severely suppressed by some fundamental mass scale and focusing in simple power counting renormalizable arguments, it turns out that Elko spinor fields may have quartic self-interaction and an Elko-Higgs (doublet) interaction\footnote{We shall emphasize that Elko does not carry standard $U(1)$ gauge invariance \cite{elko}.}. In this vein, such spinor field may act as a dark matter candidate.

Another interesting feature about Elko is its non-locality. Elko spinor fields do not belongs to a standard Wigner's class \cite{WIG}. It was demonstrated, however, that Elko breaks Lorentz symmetry (in a subtle way) by containing a preferred direction \cite{ALS}. It is worth to note that the existence of a preferred direction -- the so-called `axis of evil' -- (as well as a self interaction) is believed to be a property of dark matter \cite{MAG}. We also remark, for completeness, that the quantum field associated to the Elko spinor is now better understood in the scope of Very Special Relativity (VSR) framework \cite{VSR}. In fact, it is possible to describe, or construct, Elko spinor fields as the spinor representation of $SIM(2)$ subgroup of VSR \cite{SAL}. In this vein, since $SIM(2)$ is the largest subgroup of VSR encompassing all the necessary physical symmetries except some (violated) discrete symmetry, the tension between Elko and Lorentz symmetries disappears.   

On the other hand, it is well known that accelerators will test, in a incontestable way, theories in the scope of physics beyond the Standard Model as well as shed some light to the mass generation problem\cite{Nath:2010zj,ATLAS,Martin:1997ns,AguilarSaavedra:2005pw}. Candidates of dark matter predicted in particle physics theories, like supersymmetry, are on the focus of such studies and the answers will provide additional information for a deeper level of 
our understanding on astrophysics and cosmology. In such way, the CERN Large Hadron Collider (LHC) results are fundamental for any study connecting high energy physics and astrophysics/cosmology.  
The LHC will provide center-of-mass energy enough to probe directly the weak scale and the origin of mass. Therefore, since we still have  the open question of the dark matter nature, it is possible the study of the origin of mass as well as the candidate to the dark matter in the search of Elko. 
In considering some specific process for Elko production, radiative corrections must be taken into account. In this case, as we will see, the Elko non-locality is manifest leading to an exclusive output in the final signature.  
At phenomenological grounds, such a behavior suggests a different analysis for the search of Elko at accelerators.  So, we consider in some detail a tree level process (where the non-locality is absent) concerning to the Elko production at the LHC, whose signature is $ \mu^{+}+\mu^{-}+2\varsigma$. Such process includes the quartic self-interaction and a coupling with the Higgs scalar field. 


This paper is organized as follows: In the next Section we introduce some formal aspects of the Elko spinor fields calling attention to the main characteristics that will be relevant in the subsequent analysis. In the Section III we explore the Elko non-locality, when considering radiative corrections. In the Section IV we analyze the tree level case of a viable cross-section for Elko production at the LHC. Then, we move forward investigating some peculiar aspects of our signal. In the last Section we conclude.

\section{Elko spinor fields}

In this Section we briefly introduce the main aspects concerning the construction of Elko spinor fields. Its formal structure may be outlined as follows. Let $C$ be the charge conjugation operator given, in Weyl realization, by

\begin{eqnarray}
C=\left(
    \begin{array}{cc}
     0 & \sigma_{2} \\
      -\sigma_{2} & 0 \\
    \end{array}
  \right)K, \label{e1}
\end{eqnarray} being $K$ the operator that complex conjugate a spinor which appears on its right and $\sigma_{2}$ the usual Pauli matrix. The Elko spinor, $\lambda(\bf{p})$, is defined by
\begin{eqnarray}
C\lambda{(\bf{p})}=\pm \lambda(\bf{p}),\label{e2}
\end{eqnarray} where plus sign yields self-conjugate spinors $\big(\lambda^{S}(\bf{p})\big)$ and minus anti self-conjugate spinors $\big(\lambda^{A}(\bf{p})\big)$.
\begin{eqnarray}
\lambda(\bf{p})=\left(
                  \begin{array}{c}
                    \pm \sigma_{2} \phi_{L}^{\ast}(\bf{p}) \\
                    \phi_{L}(\bf{p}) \\
                  \end{array}
                \right).\label{e3}
\end{eqnarray} In the above equation $\phi_{L}(\bf{p})$ transform as a left handed (Weyl) spinor, hence $\sigma_{2} \phi_{L}^{\ast}(\bf{p})$ transform as a right handed spinor. In this vein, Elko spinor belongs to the $(\frac{1}{2},0)\oplus(0,\frac{1}{2})$ representation space. Now, let us set the explicit form of Elko, in the rest frame\footnote{Of course, the explicit form for any momentum is obtained by performing a boost in $\lambda(\bf{p})$.} $(\bf{p}=\bf{0})$. In order achieve the formal profile of Elko, one may look at the helicity equation $(\sigma\cdot\hat{\bf{p}})\phi^{\pm}(\bf{0})=\pm \phi^{\pm}(\bf{0})$. Taking $\hat{\bf{p}}=(sin\theta \, cos\phi, sin\theta \, sin\phi, cos\theta)$ we arrive at four spinors, following the standard notation, given by
\begin{eqnarray}
\lambda^{S}_{\{+,-\}}(\bf{0})=\left(
                              \begin{array}{cc}
                                +\sigma_{2}[\phi_{L}^{-}(\bf{0})]^{\ast} \\
                                \phi_{L}^{-}(\bf{0}) \\
                              \end{array}
                            \right)\nonumber \\
                            \lambda^{S}_{\{-,+\}}(\bf{0})=\left(
                                                    \begin{array}{cc}
                                                      +\sigma_{2}[\phi_{L}^{+}(\bf{0})]^{\ast} \\
                                                      \phi_{L}^{+}(\bf{0}) \\
                                                    \end{array}
                                                  \right)\nonumber\\
                                                  \lambda^{A}_{\{+,-\}}(\bf{0})=\left(
                              \begin{array}{cc}
                                -\sigma_{2}[\phi_{L}^{-}(\bf{0})]^{\ast} \\
                                \phi_{L}^{-}(\bf{0}) \\
                              \end{array}
                            \right)\nonumber \\
                            \lambda^{A}_{\{-,+\}}(\bf{0})=\left(
                              \begin{array}{cc}
                                -\sigma_{2}[\phi_{L}^{+}(\bf{0})]^{\ast} \\
                                \phi_{L}^{+}(\bf{0}) \\
                              \end{array}
                            \right),\label{e4}
\end{eqnarray} with phases adopted such that
\begin{eqnarray}
\phi^{+}_{L}(\bf{0})=\sqrt{m_{\varsigma}}\left(
                           \begin{array}{c}
                             cos(\theta/2)e^{-i\phi/2} \\
                             sin(\theta/2)e^{i\phi/2} \\
                           \end{array}
                         \right)\label{e5}
\end{eqnarray} and
\begin{eqnarray}
\phi^{-}_{L}(\bf{0})=\sqrt{m_{\varsigma}}\left(
                           \begin{array}{c}
                             -sin(\theta/2)e^{-i\phi/2} \\
                              cos(\theta/2)e^{i\phi/2} \\
                           \end{array}
                         \right).\label{e6}
\end{eqnarray} We remark that $-i\sigma_{2}[\phi^{\pm}_{L}(\bf{0})]^{\ast}$ and $\phi^{\pm}_{L}(\bf{0})$ present opposite helicities and, hence, Elko carries {\it both} helicities. Another important formal aspect of Elko spinor fields is its dual spinor. In order to guarantee an invariant real norm, being positive definite for two Elko spinor fields and negative definite norm for the other two, the dual for Elko is defined by
\begin{eqnarray}
{\stackrel{\neg}{\lambda}_{\{\mp,\pm\}}^{S/A}}(\bf{p})=\pm i\Big[\lambda_{\{\pm,\mp\}}^{S/A}(\bf(0))\Big]^{\dagger}\gamma^{0}.\label{e7}
\end{eqnarray} With such a definition for the Elko dual, one arrives at the following spin sums \cite{elko}
\begin{eqnarray}
\sum_{\kappa}\lambda_{\kappa}^{S}{\stackrel{\neg}{\lambda}_{\kappa}^{S}}=+m_{\varsigma}[\mathbb{I}+\mathcal{G}(\phi)]\nonumber\\
\sum_{\kappa}\lambda_{\kappa}^{A}{\stackrel{\neg}{\lambda}_{\kappa}^{A}}=-m_{\varsigma}[\mathbb{I}-\mathcal{G}(\phi)],\label{e8}
\end{eqnarray} where $\mathcal{G}(\phi)$ is given by \cite{ALS}
\begin{eqnarray}
\mathcal{G}(\phi)=\gamma^{5}(\gamma_{1}sin\phi-\gamma_{2}cos\phi),\label{e9}
\end{eqnarray} and the gamma matrices are
\begin{eqnarray}
\gamma^{0}=\left(
             \begin{array}{cc}
               0 & 1 \\
               1 & 0 \\
             \end{array}
           \right),\label{e10} \hspace{2cm} \gamma^{i}=\left(
                                   \begin{array}{cc}
                                     0 & -\sigma^{i} \\
                                     \sigma^{i} & 0 \\
                                   \end{array}
                                 \right), \label{e11}
\end{eqnarray} being $\gamma^{5}=-i\gamma^{0}\gamma^{1}\gamma^{2}\gamma^{3}$. Spin sums entering in a profound level into the local structure, as well as the statistic, of the theory. It is important to note that the right-hand side of Eqs. (\ref{e8}) is not proportional (or unitary connected) to the momentum operators\footnote{In acute contrast with the usual Dirac case.}. Therefore the relations (\ref{e8}) are responsible for the peculiar characteristics of Elko locality structure, as well as its breaking of Lorentz invariance. Such peculiarity, obviously, brings important modifications in the S-matrix calculations (see next Section). 

After studying the formal structure of Elko spinor fields, we shall examine the quantum field associated to such spinor. It is possible to define an Elko-based quantum field, respecting its formal properties, by
\begin{eqnarray}
\eta(x)=\int \frac{d^{3}p}{(2\pi)^{3}}\frac{1}{\sqrt{2mE(\bf{p})}}\sum_{\alpha}[c_{\alpha}(\bf{p})\lambda^{S}_{\alpha}(\bf{p})
e^{-ip_{\mu}x^{\mu}}+c_{\alpha}^{\dagger}(\bf{p})\lambda^{A}_{\alpha}e^{+ip_{\mu}x^{\mu}}],\label{e12}
\end{eqnarray} being $c_{\alpha}^{\dagger}(\bf{p})$ and $c_{\alpha}(\bf{p})$ the creation and annihilation operators, respectively, satisfying the fermionic anticommutation relations
\begin{eqnarray}
\{c_{\alpha}(\bf{p}),c_{\alpha'}^{\dagger}(\bf{p'})\}=(2\pi)^{3}\delta^{3}(\bf{p}-\bf{p'})\delta_{\alpha\alpha'},\\
\{c_{\alpha}^{\dagger}(\bf{p}),c_{\alpha'}^{\dagger}(\bf{p'})\}=\{c_{\alpha}(\bf{p}),c_{\alpha'}(\bf{p'})\}= 0.\label{e13}
\end{eqnarray} The Elko dual $\stackrel{\neg}{\eta}$ is obtained by replacing $\lambda$ by its dual, $c$ by $c^{\dagger}$ and $ip_{\mu}x^{\mu}$ by $-ip_{\mu}x^{\mu}$ (and vice-versa). There is a crucial identity obeyed by Elko, given by the application of the $\gamma_{\mu}p^{\mu}$ operator to $\lambda^{S/A}(\bf{p})$:
\begin{eqnarray}
(\gamma_{\mu}p^{\mu}\delta_{\alpha}^{\beta}\pm im \mathbb{I}\varepsilon_{\alpha}^{\beta})\lambda_{\beta}^{S/A}(\bf{p})=0,\label{e14}
\end{eqnarray} where $\varepsilon_{\{+,-\}}^{\{-,+\}}:=-1$ and $\delta_{\alpha}^{\beta}$ is the usual Kronecker symbol. In view of (the simply algebraic) Eq. (\ref{e14}) it turns out that Elko satisfy the Klein-Gordon (not Dirac) equation and, therefore, it must be associated to a Klein-Gordon-like Lagrangian:

\begin{eqnarray}
\mathcal{L}^{free}=\partial^{\mu}\stackrel{\neg}{\eta}(x)\partial_{\mu}\eta(x)-m^{2}_{\varsigma}\stackrel{\neg}{\eta}(x)\eta(x).\label{EI}
\end{eqnarray}

As already mentioned in the Introduction, we shall study the coupling between Elko and Higgs 
fields, since it is the unique renormalizable (perturbatively) Elko coupling. Therefore, in the next Section we shall explore the features of the (\ref{EI}) lagrangian, plus the interaction given by 

\begin{eqnarray}
\mathcal{L}^{int}= \lambda_{\varsigma}\phi^{2}(x)\stackrel{\neg}{\eta}(x)\eta(x).\label{TA}
\end{eqnarray}  

In this work, and consequently to obtain the Feynman rules relevant to it (see Ref.\cite{diagrammar}), our object of study is (\ref{EI}) and (\ref{TA}) added with the usual kinetic and interaction terms for the Higgs boson,  the Z vector field and summing over all the quarks in the theory, as they appear in the Standard Model after symmetry breaking. 

As a last remark we emphasize that, in general, Eqs. (\ref{e8}) and (\ref{e9}) suggest that there is a preferred axis for Elko. In fact, it is possible to show that Elko enjoy locality in the direction perpendicular to its plane \cite{ALS}, or, equivalently, along the preferred axis $\hat{z}_{e}$. Let us give an example coming from the canonical structure of Elko fields in order to clarify this point. The canonical conjugate momenta to the Elko fields are given by
\begin{eqnarray}
\Pi(x)=\frac{\partial \mathcal{L}_{KG}}{\partial \dot{\eta}}=\frac{\partial \stackrel{\neg}{\eta}}{\partial t},\label{e15}
\end{eqnarray} where $\mathcal{L}_{KG}$ stands for a Klein-Gordon-like Lagrangian. The equal time anticommutator for $\eta(x)$ and its conjugate momentum is
\begin{eqnarray}
\{\eta(\bf{x},t),\Pi(\bf{x'},t)\}=i\int\frac{d^{3}p}{(2\pi)^{3}}\frac{1}{2m}e^{i\bf{p}\cdot(\bf{x}-\bf{x'})}\sum_{\alpha}
\Big[\lambda_{\alpha}^{S}(\bf{p})\stackrel{\neg}{\lambda}_{\alpha}^{S}(\bf{p})-\lambda_{\alpha}^{A}(\bf{-p})
\stackrel{\neg}{\lambda}_{\alpha}^{A}(\bf{-p})\Big],\label{e16}
\end{eqnarray} which, in the light of the spin sums, may be recast in the following form
\begin{eqnarray}
\{\eta(\bf{x},t),\Pi(\bf{x'},t)\}=i\delta^{3}(\bf{x}-\bf{x'})\mathbb{I}+i\int \frac{d^{3}p}{(2\pi)^{3}e^{i\bf{p}\cdot(\bf{x}-\bf{x'})}}\mathcal{G}.\label{e17}
\end{eqnarray} The existence of a preferred axis is now evident, since the second integral in the right-hand side Eq. (\ref{e17}) vanishes along the $\hat{z}_{e}$. So, this preferred axis may be understood as an axis of locality.

\section{  Exploring  Elko non-locality }

\begin{figure}[h]
\begin{center}
\includegraphics[scale=.4]{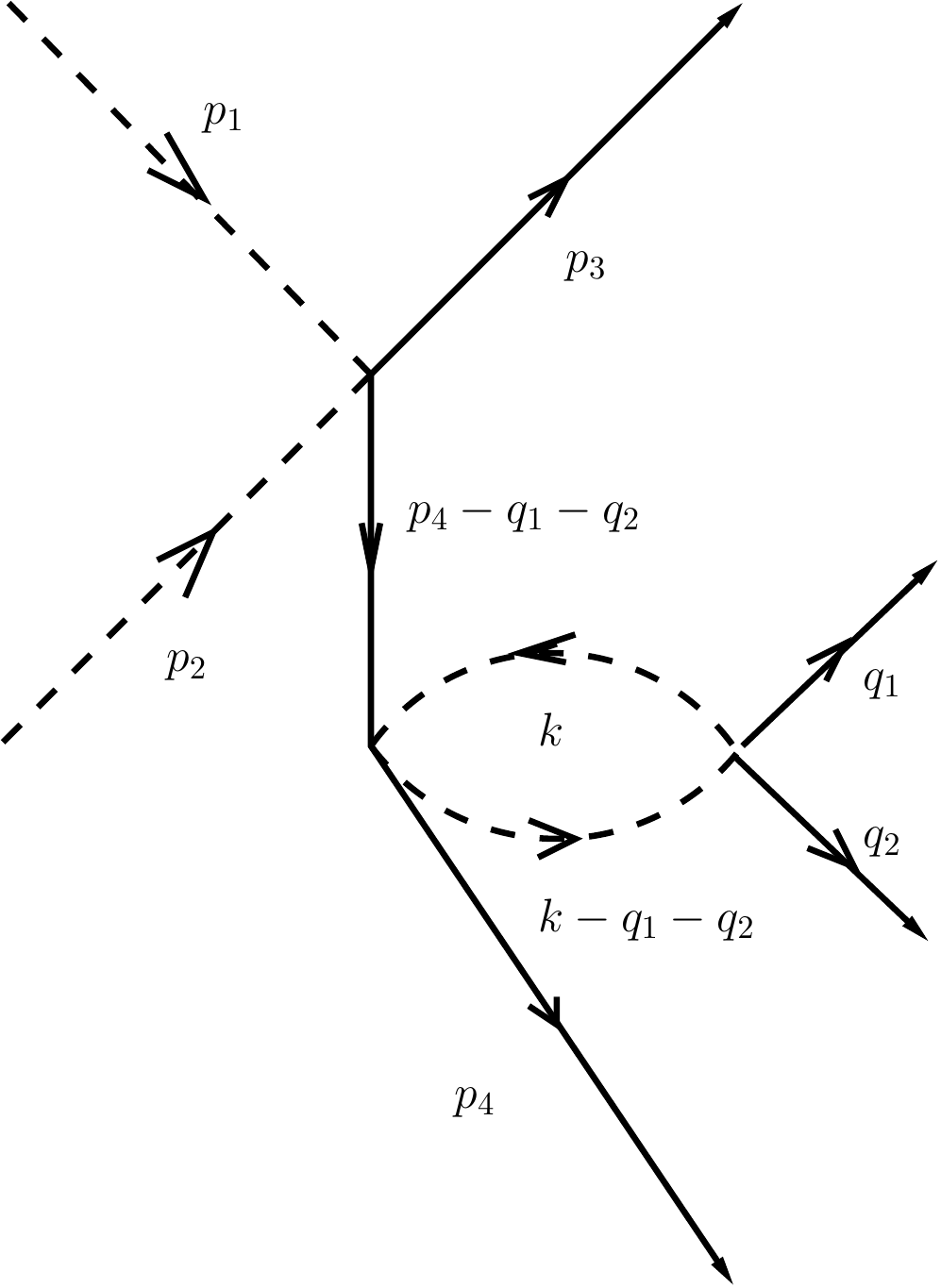}
\caption{Example of higher order graphic relevant to Elko production and its non locality. Dotted lines stands for Higgs boson and continuous lines for Elko. }\label{evile5}
\end{center}
\end{figure}

According to its typical Lagrangian Elko spinor fields couples only to the Higgs boson and, hence, any production mechanism of such particle must occur via Higgs production or decay process. 
A very specific feature of Elko production is its non locality, encoded in the propagator behavior which has a different form (the $\mathcal{G}(\phi)$ term appears explicitly)  when  computed outside its axis of propagation. In order to explore a little further this effect, let us consider for instance the first graph of a cascade production of Elko particles (Fig. (\ref{evile5})).

If one chooses to compute (or measure) such a higher order process in the same plane where the intermediary Elko is propagating, the amplitude reads
\begin{eqnarray}\nonumber
i\cal{M}&=&\lambda_{\varsigma}^3\frac{\lambda^A_\alpha(p_3)\lambda^A_\rho(q_1)\stackrel{\neg}{\lambda}_{\beta}^{S}(p_4)\stackrel{\neg}{\lambda}_{\sigma}^{S}(q_2)}{(p_4+q_1+q_2)^2-m_{\varsigma}^2}\int\frac{d^4k}{(2\pi)^4}\frac{1}{[k^2-m^2_H][(k-q_1-q_2)^2-m_H^2]}.
\end{eqnarray}
Otherwise, there is also in the amplitude the presence of the $\mathcal{G}(\phi)$ term
\begin{eqnarray}\nonumber
i\cal{M}&=&\lambda_{\varsigma}^3\frac{\lambda^A_\alpha(p_3)\lambda^A_\rho(q_1)[1+\mathcal{G}(\phi)]\stackrel{\neg}{\lambda}_{\beta}^{S}(p_4)\stackrel{\neg}{\lambda}_{\sigma}^{S}(q_2)}{(p_4+q_1+q_2)^2-m_{\varsigma}^2}\int\frac{d^4k}{(2\pi)^4}\frac{1}{[k^2-m^2_H][(k-q_1-q_2)^2-m_H^2]}.
\end{eqnarray} 
The divergence appearing in the above amplitude was treated via Pauli-Villars regularization, subtracted this amplitude from its value at $q_1=q_2=0$. The result is given by
\begin{eqnarray}
i{\cal{M}}_{RG}&=&\lambda_{\varsigma}^3\frac{\lambda^A_\alpha(p_3)\lambda^A_\rho(q_1)[\mathbb{I}+\mathcal{G}(\phi)]\stackrel{\neg}{\lambda}_{\beta}^{S}(p_4)\stackrel{\neg}{\lambda}_{\sigma}^{S}(q_2)}{(p_4+q_1+q_2)^2-m_{\varsigma}^2}\int_0^1ln\left(\frac{(q_1+q_2)^2x(x-1)+m_H^2}{m_H^2}\right).
\end{eqnarray}
Computing the traces (where $E_1$ and $E_2$ are, respectively $q_1$ and $q_2$ particle energies) the average spin squared sum is 
\begin{eqnarray}\nonumber
\frac{1}{16}\sum_{spins}|{\cal{M}}_{RG}|^2&=&\frac{E_2E_4(E_3+p_3)(E_1+q_1)trace\left[(\mathbb{I}-\mathcal{G}(\phi))(\mathbb{I}+\mathcal{G}(\phi))(\mathbb{I}+\mathcal{G}(\phi))\right]trace\left[\mathbb{I}-\mathcal{G}(\phi)\right]}{[(p_4+q_1+q_2)^2-m_{\varsigma}^2]^2}\nonumber\\
&\times&\left[\int_0^1ln\left(\frac{(q_1+q_2)^2x(x-1)+m_H^2}{m_H^2}\right)\right]^2\lambda_{\varsigma}^6\nonumber\\
&=&\lambda_{\varsigma}^6\frac{8E_2E_4(E_3+p_3)(E_1+q_1)}{[(p_4+q_1+q_2)^2-m_{\varsigma}^2]^2}\left[\int_0^1ln\left(\frac{(q_1+q_2)^2x(x-1)+m_H^2}{m_H^2}\right)\right]^2.
\end{eqnarray}
Note that if one lies on the $\vec{p}_4+\vec{q}_1+\vec{q}_2$ direction the obtained result is divided by two. Since the decay rate is proportional to the average spin squared amplitude integrated over the four-body phase space, the Elko particle decay in a preferred axis. Besides, the decay process in such a channel is one half lower than in any other direction.

The above considerations lead to an important result: if the cut applied on $\phi$ includes the intermediary Elko propagation axis, the measured decay is lower than any other cut in which this specific direction is not included. Therefore it breaks $\phi$ isotropy which is, obviously, fully observed in all Standard Model particles. Such a process makes then manifest the Elko non locality, giving also a clue for its signature. 
\begin{figure}[t]
\begin{center}
\includegraphics[scale=.6]{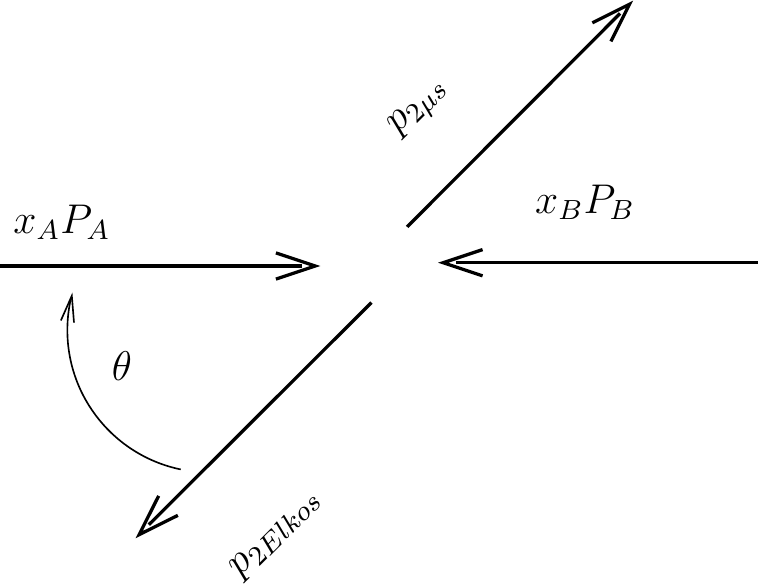}
\caption{Kinematics of Elko production. }\label{evile6}
\end{center}
\end{figure}
We should also note another feature in this production, as reflect of momentum conservation,  represented in Fig. (\ref{evile6}). An increase in the Elko production, in a preferred direction, should implicate a decrease of the remain particles
final momentum in the same direction (as a missing energy in the detector), reflecting in a complementary angular distribution, when compared with its   possible {\it background}.

\section{Tree level case }

For tree level calculations, the non-locality effect is not manifest, and the study of possible signals of Elko decay at accelerators is addressed to the standard searching. For this purposes, we have considered the case where 
Elko can be produced at the LHC through the Higgs boson fusion, via quartic coupling as depicted in Fig. (\ref{anthrax1}). In both cases (Higgs production or decay process), however, the production is suppressed according to the value of the coupling constant, leaving the number of events and the signature of the decay expressed as a function of two fundamental parameters of the model: the Elko mass and the Elko-Higgs boson coupling constant, which will be taken as less than or equal to one, in order to ensure renormalizability. At the LHC, signatures with leptons as a final state are preferred, specially muons, whose background can be calculated directly from the Standard Model. Besides, the identification of muons are well given as, for example, at CMS technical proposal. In this vein, we will be focused in a two muons + Elko signal, according to the process illustrated in the graph (Fig. (\ref{anthrax1})). In this case the process is  $q+\bar{q}\to \mu^+ +\mu^- +2\varsigma$ , where $2\varsigma$ stands for the two Elko particles with mass $m_{\varsigma}$ produced in the threshold were they will be on rest in the CoM frame. 
We do not considered here the direct production of two Higgs from Elko fusion, since the Higgs boson is, indeed, the key block to be detected at the LHC. We have fixed the Higgs mass boson in the experimental limit~\cite{CDFNote} and also considered jets with high energy and momentum. In such case, they will emerge almost collinear with the beam. The interaction rate is proportional to the cross section calculated as follow:
\begin{figure}[htb]
\centering
\includegraphics[scale=.3]{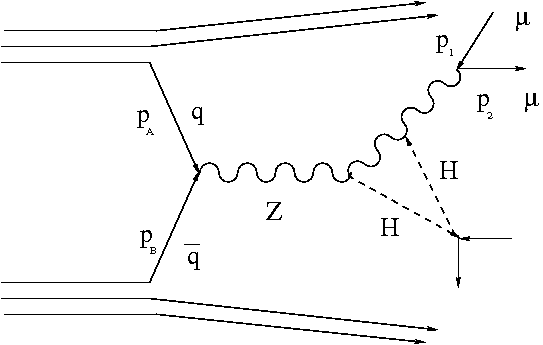}
\caption{ $q+\bar{q}\to \mu^+ +\mu^- +2\varsigma$ scattering. The loop is composed by  two Higgs and a Z boson. }\label{anthrax1}
\end{figure}

We shall label $p_A=x_AP_A$ and $p_B=x_BP_B$, respectively, as the momentum for the quark and anti-quark, related to the initial protons $P_{A,B}$ and the muons with momentum $p_1$ and $p_2$. The amplitude is given by:
\begin{eqnarray}\label{anthrax0}\nonumber
i{\cal M}&=& q^r(p_A)\left[\frac{ig_Z}{2}\gamma^\mu (c_V^f-c_A^f\gamma^5)\right]\bar{q}^{r'}(p_B)\left[\frac{-i}{q^2-m^2_Z}\left(g_{\mu\nu}-\frac{q_\mu q_\nu}{m_Z^2}\right)\right]\frac{igm_Zg^{\nu\rho}}{2\cos{\left(\theta_w\right)}}\left[\frac{-i}{k^2-m^2_Z}\left(g_{\rho\sigma}-\frac{k_\rho k_\sigma}{m_Z^2}\right)\right.\\\nonumber
&\times&\left.\frac{i}{(q-k)^2-m^2_H}\right]\frac{igm_Z\, g^{\sigma\gamma}}{2\, \cos{\left(\theta_w\right)}}\left[\frac{i}{(q-k)-m_H^2}\right]\lambda_{\varsigma}{\stackrel{\neg}{\lambda}^{S}}_\Lambda{\lambda}^{A}_\Omega\left[\frac{-i}{q^2-m_Z^2}\left(g_{\gamma\delta}-\frac{q_\gamma q_\delta}{m_Z^2}\right)\right]\\
&\times& \frac{-ig_Z}{2}\gamma^\delta\left(-\frac{1}{2}+2\, \sin^2{\left(\theta_w\right)}+\frac{1}{2}\gamma^5\right)\bar{u}^s(p_1)v^{s'}(p_2),
\end{eqnarray}
following the conventions of Ref. \cite{diagrammar}, where the factors for quarks reads
\begin{eqnarray*}
u&\Rightarrow& c_A^f=1/2, \;\;\; c_V^f=1/2 -4/3 \sin^2{(\theta_w)}\\\nonumber
d&\Rightarrow& c_A^f=-1/2, \;\;\; c_V^f=-1/2 +2/3 \sin^2{(\theta_w)}.
\end{eqnarray*}
On partonic CoM reference frame and $p_A=p_B=p_1=p_2\approx 0$ we can set
\begin{eqnarray*}
p_A&=&\frac{\sqrt{\hat{s}}}{2}(1,0,0,1), \;\;\; p_B=\frac{\sqrt{\hat{s}}}{2}(1,0,0,-1)\\\nonumber
p_1&=&\left(\frac{\sqrt{\hat{s}}}{2}-m_{\varsigma}\right)(1,\sin{(\theta)},0,\cos{(\theta)}),\;\;  p_2=\left(\frac{\sqrt{\hat{s}}}{2}-m_{\varsigma}\right)(1,-\sin{(\theta)},0,-\cos{(\theta)})\\\nonumber
p_3&=&m_{\varsigma}(1,0,0,0),\;\;\;\;\; p_4=m_{\varsigma}(1,0,0,0),
\end{eqnarray*}
where $q=\sqrt{\hat{s}}$.

\begin{figure}
\centering
\includegraphics[scale=.3]{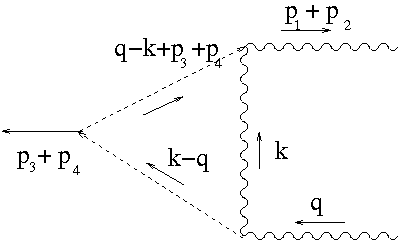}
\caption{Performed loop calculation.}\label{anthrax3}
\end{figure}

Looking at Fig. (\ref{anthrax3}) we can identify

\begin{eqnarray}\label{anthrax6}
P_1&=& (k-q)^2-m_H^2\\\nonumber
P_2&=& (q+2m_{\varsigma}-k)^2-m_H^2\\\nonumber
P_3&=& k^2-m_Z^2=l_0^2-l_\perp^2-m_Z^2,
\end{eqnarray}
as the denominators for the function to be integrated. In order to use the functions well established (\verb+OneLoop2Pt+) on \verb+xloops+ package \cite{xloops} we need to reduce the number of functions on denominator  (\ref{anthrax6}), using Feynman trick,
\begin{eqnarray}\label{anthrax7}
\frac{1}{P_1P_2P_3}&=&\int_0^1\frac{1}{P_3}\frac{dx}{[P_1x+P_2(1-x)]^2}=\int_0^1\frac{1}{P_3}\frac{dx}{(k+q^\prime)^2-m^2}\\\nonumber
q^\prime&=&-x\sqrt{\hat{s}}+(x-1)(\sqrt{\hat{s}}+2m_{\varsigma})\Rightarrow e_0^\mu=\frac{q^{\prime\mu}}{||q^\prime||}=-(1,0,0,0)\\\nonumber
m^2&=&[x\sqrt{\hat{s}}+(1-x)(\sqrt{\hat{s}}+2m_{\varsigma})]+m_H^2-x\sqrt{\hat{s}}-(1-x)(\sqrt{\hat{s}}+2m_{\varsigma})^2,
\end{eqnarray}
where $x$ integration was performed with \verb+Maple+ using of the approximation where $m_{\varsigma}/\sqrt{\hat{s}}\approx 0$.  Obviously, such an approximation in the Elko mass is largely justified in order to guarantee Elko spinor fields as a dark matter candidate. This choice restrict the experimental analysis to events with low energy QCD jets in its final state, since almost all momentum is transferred to the initial partons, providing a signature for the Elko production. One can expect a missing energy on detectors, due to the fact that Elko particles will be unobserved by detectors and the only impact in its production is reduce the final $\mu^+ +\mu^-$ quadrimomentum.
With this expression at hands, it is necessary to multiply  by its conjugate and perform the respective polarization sums (\ref{e8}), taking into account, obviously, the terms $\mathcal{G}(\phi)$ responsible for the non-locality 
outside $\hat{z}$ axis. Is straightforward to perform those traces for Elko polarization sums using the Elko dual definitions and the spin sums \cite{elko}
\begin{eqnarray}
\sum_{\kappa}\stackrel{\neg}{\lambda}_{\kappa}^{S}\left(\stackrel{\neg}{\lambda}_{\kappa}^{S}\right)^\dagger&=&\sum_{\kappa}\left(i\epsilon_\kappa^\rho{\lambda^S_\rho}^\dagger\gamma^0\right)\left(i\epsilon_\kappa^\sigma{\lambda_\sigma^S}^\dagger\gamma^0\right)^\dagger=\sum_\kappa \epsilon_\kappa^\rho\epsilon^\sigma_\kappa {\lambda^S_\rho}^\dagger\lambda_\sigma^S\\
&=&{\lambda^S_{\{-,+\}}} ^\dagger\lambda^S_{\{-,+\}}+{\lambda^S_{\{+,-\}}} ^\dagger\lambda^S_{\{+,-\}}=4E\mathbb{I},
\end{eqnarray}
where $\epsilon^{\{-,+\}}_{\{+,-\}}=-\epsilon^{\{+,-\}}_{\{-,+\}}=-1$.

After squaring, taking traces and averaging over the spin of the initial and final particles (we approximate the masses for quarks and muons to zero), we should obtain $\sum_{r,r'} \sum_{s,s',\Omega,\Lambda}|{\cal M}|^2$. One could use it to calculate
\[d\hat{\sigma}=\frac{1}{2E_A2E_B}\frac{1}{2}\left(\frac{1}{64}\sum_{spin}|{\cal M}|^2\right)\;dPS,\]
where $dPS$ is the phase space for two muons with momentum $p_1$ and $p_2$ and two Elkos with mass $m_\varsigma$ on rest, i. e., 
\begin{eqnarray}
dPS&=&(2\pi)^4\delta^4(p_A+p_B-p_1-p_2-p_3-p_4)\frac{d^3p_1}{(2\pi)^3(2E_1)}\frac{d^3p_2}{(2\pi)^3(2E_2)}\\\nonumber
&=&\frac{1}{4(2\pi)^2}\delta(\sqrt{\hat{s}}-E_1-E_2-2m_{\varsigma})\frac{p_1^2dp_1\, d\Omega}{E_1E_2}=\frac{1}{32\pi^2}\frac{\sqrt{\hat{s}}}{\sqrt{\hat{s}}-2m_{\varsigma}}d\Omega,
\end{eqnarray}
where $|p_1|\, dp_1=E_1\, dE_1$, being $E_1=\left(\sqrt{\hat{s}}/2-m_{\varsigma}\right)$. We emphasize that we are working within  $m_{\varsigma}\approx 0$ approximation. We also stress that 
 $d\hat{\sigma}$ has no dependence on angular coordinates, so the integration on $d\Omega$ gives a multiplicative factor $4\pi$ for the total cross section. Our final result, however, is to much huge to be presented here.



\begin{figure}[hbt]
\begin{minipage}[t]{0.4\linewidth}
\begin{center}
\includegraphics[scale=.38]{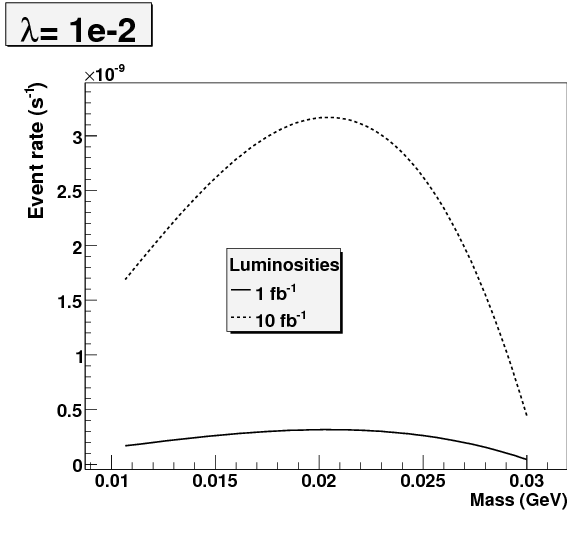}
\end{center}
\centering{(a)}
\end{minipage}
\hspace{0.9cm}
\begin{minipage}[t]{0.4\linewidth}
\begin{center}
\includegraphics[scale=.38]{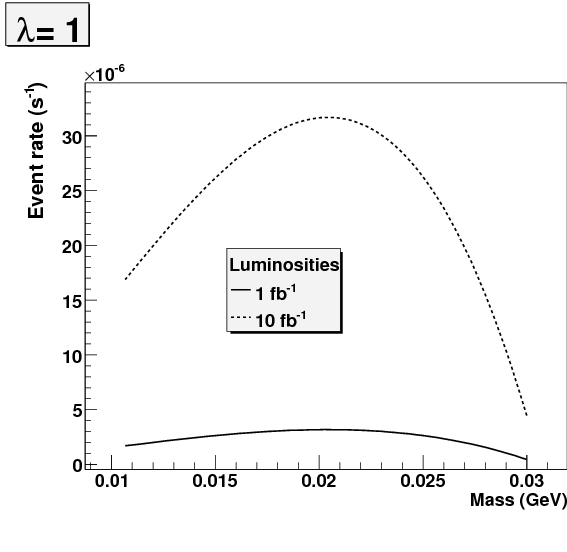}
\centering{(b)}
\end{center}
\end{minipage}
\begin{minipage}[t]{0.4\linewidth}
\begin{center}
\includegraphics[scale=.38]{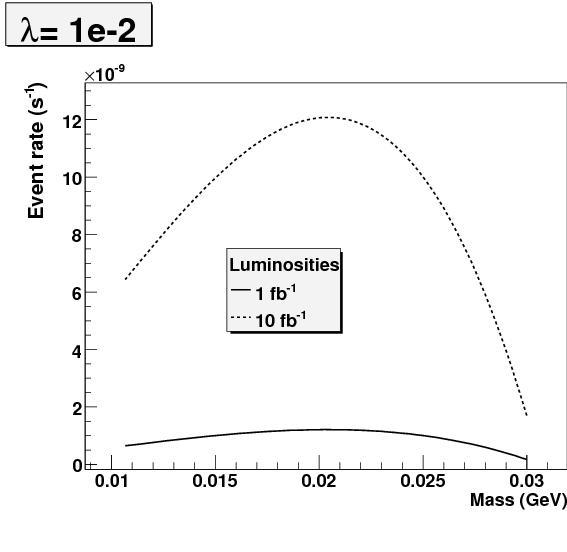}
\centering{(c)}
\end{center}
\end{minipage}
\hspace{0.9cm}
\begin{minipage}[t]{0.4\linewidth}
\begin{center}
\includegraphics[scale=.38]{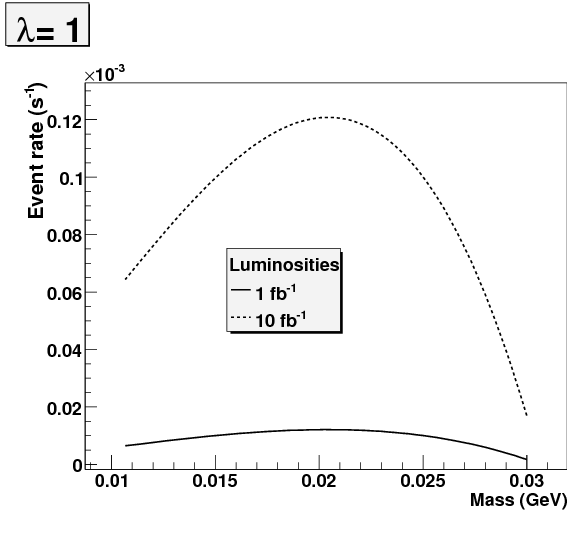}
\centering{(d)}
\end{center}
\end{minipage}

\caption{Event rate $(1/s)$ versus mass ($GeV$) for two luminosity values and the center-of-mass energy at the LHC for 7 TeV (a-b) and 14 TeV (c-d). The range for mass was chosen to guarantee the fact that the Elko can be a possible candidate for dark matter \cite{elko}. We have considered two values for $\lambda_{\varsigma}$, namely, $1$ and $10^{-2}$.}\label{anthrax15}
\end{figure}

On the hadronic frame, 
$P_A=\frac{\sqrt{s}}{2}(1,0,0,1)$ and $P_A=\frac{\sqrt{s}}{2}(1,0,0,-1)$. Thus
\begin{eqnarray*}
s=(P_A+P_B)^2=\frac{\hat{s}}{x_A x_B},
\end{eqnarray*}
and we will integrate using \verb+Cuba+ routines \cite{cuba}
 \begin{eqnarray*}
\sigma_{(p+p\to \mu^++\mu^-+2\varsigma)}=\sum_q\int_0^1\int_0^1 dx_A dx_B[f_q(x_A)f_{\bar{q}}(x_B)+f_{\bar{q}}(x_A)f_q(x_B)] \hat{\sigma}({\hat{s}}) \delta(\hat{s}-x_A x_B s).
\end{eqnarray*}
With the hadronic total cross section at hands, it is straightforward to obtain the event rate $R$ by multiplying $\sigma$ by the integrated luminosity ${\cal L}$, estimated in $1 fb^{-1}$ and $10 fb^{-1}$. 

The results of the studied process are presented in Fig. (\ref{anthrax15}). We show the total expected event rate for 2 Elkos + $\mu^{+} \mu{-}$ via the Higgs boson fusion, at the LHC, for two different values of the center-of-mass energy, as well as two different values for the total luminosity. The total number of events is presented as a function of the Elko mass. The main case we consider, with total luminosity of $10fb^{-1}$, at 7 TeV, for a coupling constant of an order of 1 shows a quite optimistic number of events, around a thousand. For a smaller coupling constant, $O$$(10^{-2})$, the number of events is also large. In this sense, we can consider the LHC, for instance, as a good scenario to study both, the Higgs boson and the Elko production in order to shed some light to the dark matter problem. For a 14 TeV center-of-mass energy case, in both $1fb^{-1}$ and $10fb^{-1}$ cases, the total number of events produced at the LHC is even bigger, for the different values of the coupling constant. By now, since the number of events is encouraging, we shall keep our attention in the exploration of a typical signature encoding the Elko non-locality. 

\section{Detection possibility at LHC}

Even though the decay in the preferred axis is estimated as one half lower than in any other direction, a poor detector angular resolution on this decay will smear out this effect, either due to the detector tracking, or due to the poor event reconstruction. Therefore it is mandatory to make an estimation of the minimum angular resolution requirement to detect this effect. At the LHC, the minimum angular resolution at, e.g., the  CMS detector $\Delta\phi_{res}=10\, mrad$ \cite{:1994pu}. The relative significance on this interval for an integrated luminosity $L$, taking into account our background will be given by
\begin{eqnarray}\label{artill}
S_{rel}=\frac{S}{\sqrt{B}},
\end{eqnarray} since the background is isotropically distributed in the azimuthal angle and the efficiency on the muon measurement is about $98\%$. In Eq. (\ref{artill}), $S$ stands for the number of events produced in the Elko decay and $B$ denotes the number of events related to the background.

The signal is characterized by a dimuon in the final state reconstructed in a $Z$ boson and some missing energy in the final state. Thus the irreducible SM background consists of the ZZ decaying in two muons and two neutrinos, as already studied in Ref. \cite{Cheung:2010af}. The background processes for the signal, considering next-to-leading order cross section are presented in Table  \ref{tabelasecaodechoque} (see \cite{:1994pu}).
\begin{table}
\begin{center}
\begin{tabular}{cc}
Channel & Cross Section (pb)\\\hline
$qq\rightarrow WW \rightarrow \mu^{+} \mu^{-}$& 11.7\\\hline
$t \bar{t}$&840\\\hline
$gg \rightarrow WW \rightarrow \mu^{+} \mu^{-}$ & 0.54\\\hline
$\gamma^*$ ,$ Z$&145000\\\hline
$b \bar{b} \rightarrow \mu^{+} \mu^{-}$ & 710 \\\hline
$ZW \rightarrow \mu^{+} \mu^{-} l^{\pm}$ &1.63 \\\hline
$tWb \rightarrow \mu^{+} \mu^{-}$ & 3.4 \\\hline
$ZZ \rightarrow \mu^{+} \mu^{-} $& 1.52 \\\hline
\end{tabular}
\end{center}
\caption{Background estimative for the high-order process under study.}\label{tabelasecaodechoque}
\end{table}
The irreducible SM background for the signal is the ZZ process, where one of the Z bosons 
decays into neutrinos. Since we are interested in making an estimate of the signal, taking 
into account the background without defining cuts for a detailed analysis, we only consider 
the ZZ process, as it is much larger than the signal. 
In such case, the number of events for the background considering a luminosity of $2\times 10^{33}cm^{-2}s^{-1}$, is around 99000 events. 

\begin{figure}
\centering
\includegraphics[scale=.4]{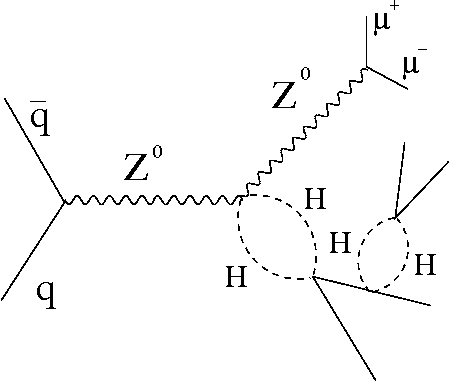}
\caption{Feynman diagram for the production of 4 missing Elko bosons (solid lines) and two muons, associated with some Higgs intermediary process.}\label{big3}
\end{figure}

In order to explore the claimed angular dependence for the signal, we study the process $q\bar{q}\to Z^{*}\varsigma\varsigma^{*}\to 2\mu 4\varsigma$, which is shown in Fig. (\ref{big3}). The two final muons inherits the sensibility on azimuthal angle by momentum conservation on the final states. Actually this process is nothing but that one described in Fig. (\ref{anthrax1}) followed by the decay of Fig. (\ref{evile5}), mediated by two loops involving Higgs particles.

An analitic expression for this process can be obtained using the equation for  $\frac{1}{16}\sum_{spins}|{\cal{M}}_{RG}|^2$ (see Section III) and  supposing the limit $\frac{q_1+q_2}{m_E}\approx 0, q_1+q_2>p_4$. One can expand the integrand, and proceed with the integration for the first term to obtain
\begin{eqnarray*}
\frac{1}{16}\sum_{spins}|{\cal{M}}_{RG}|^2\approx \lambda_\varsigma^6\frac{8E_2E_4(E_3+p_3)(E_1+q_1)}{(p_4+q_1+q_2)^4}\frac{(q_1+q_2)^4}{36m_H^4}\approx \lambda_\varsigma^2\frac{2E_2E_4(E_3+p_3)(E_1+q_1)}{9m_H^4}.
\end{eqnarray*}
In the limit $p_3,p_4\to 0$, and with all final state energies near to the Elko mass, we obtain a lower bound to this values given by
\begin{eqnarray}\label{marit}
\frac{1}{16}\sum_{spins}|{\cal{M}}_{RG}|^2\geq \lambda_\varsigma^6\frac{2m_E^4}{9m_H^4},
\end{eqnarray}
which shall be multiplied by the the cross section obtained numerically  before.

For the Elko production, on the simple decay $2\mu+2\varsigma$ and fixing the coupling constant at its maximum value ($\lambda_{\varsigma} = 1$) as well as $ m_\varsigma=0.09 GeV$, we have $\sigma_{signal}=5.06 \, fb$. At the LHC, for a $1\, fb^{-1}$ integrated luminosity, one should obtain  a  ratio $\frac{S}{\sqrt{B}}=\frac{\sigma_{signal}\sqrt{L}}{\sqrt{\sigma_{bckg}}}$ around $5$ (or actually one half of this value, taking account the angular asymmetry). 

However as  the detection of this process signature depends on the coupling constant, since under a certain value it would be required a better angular resolution in the detector to distinguish the signal from the background. 
For an indirect search of Elko particles via azimuthal angular asymmetry with $2\mu+4\varsigma$ process, using the angular resolution for the CMS detector ($\Delta\phi\geq\Delta\phi_{res}$), $m_\varsigma=0.09 \, GeV$ and $ \sqrt{s}=7\, TeV$, the number of events decreases substantially to $S=4.4\times 10^{-15}$ taking (\ref{marit}) into account. Hence, one can see that  for this parameters Eq. (\ref{artill}) gives a result which is clearly insufficient to claim a discovery at the LHC. 
The process on study has actually a   dependency on $\lambda_\varsigma^6$, so the estimated minimum resolution for the $\lambda_\varsigma=1\times 10^{-2}$ case, maintaining $S_{rel}\approx 5$, is $\Delta\phi_{res}\approx 9.1\times  10^{-11}\, rad$. Lower  values of $\lambda_\varsigma$ should require a better resolution on the detector. Of course, for this rough estimate, none type of cuts was performed and a detailed study using a Monte Carlo simulation for the final state Elko momenta would be in order. 

The main motivation for this analysis is the possibility of Elko detection in a range of parameters making possible to address Elko as possible dark matter candidate.
We now shall look at the following question: what should be the  expected missing energy in the dimuon+jet system, if Elko production is occurring taking into account the Elko non-locality? Considering the proton-proton energy as $(\sqrt{s},0,0,0)$ in Fig. \ref{evile6} we have the momentum configuration
\begin{eqnarray*}
p_{2\mu}&=&\frac{\sqrt{s}}{2}\left(1+\frac{m_{2\mu}}{s}-\frac{m_{2\varsigma}}{s},\beta sin(\theta),0,\beta cos(\theta)\right),\\
p_{2\varsigma}&=&\frac{\sqrt{s}}{2}\left(1+\frac{m_{2\varsigma}}{s}-\frac{m_{2\mu}}{s},-\beta sin(\theta),0,-\beta cos(\theta)\right),\\
\end{eqnarray*}
where $m_{2\varsigma} (m_{2\mu})$ is the invariant mass, for instance $m_{2\varsigma}=2m_{\varsigma}^2-2\vec{p}_3\cdot\vec{p}_4+2E_3E_4$, as the sum of two momentum vectors,  and $\beta=\sqrt{1-2\frac{m_{2\mu}+m_{2\varsigma}}{s}+\frac{(m_{2\mu}-m_{2\varsigma})^2}{s^2}}$. Therefore  the missing energy is
\[E^{miss}=\sqrt{s}\left(1+\frac{m_{2\mu}-m_{2\varsigma}}{s}\right)-\sqrt{s}=\frac{m_{2\mu}-m_{2\varsigma}}{\sqrt{s}}.\]

An important requirement is imposed by the minimum energy resolution for the search of missing energy on this channel. Considering the same parametrization as used  for the CMS detector \cite{:1994pu}, we suppose that the threshold for the missing energy for the signal is given by
\[E^{miss}=\frac{m_{2\mu}-m_{2\varsigma}}{\sqrt{E}}.\]
In the limit that the two elkos does not have  a significant momenta, it is possible to approximate $m_{2\mu}\approx m_Z=91.187\, GeV$ and, then, one should to select only events with  $E_{miss}>25\, GeV$. This means that a detailed analysis should take into account both, angular and energy, resolutions.

\section{Final remarks}


 By analyzing the consequences of the unusual Elko propagator behavior, it was possible to derive a typical signature to the Elko production, namely: due to the Elko non locality, the measured decay depends on the angular cut applied, breaking therefore the angular isotropy (fully observed in all standard model processes). 
 
We shall stress two important points: Fig. \ref{evile5} may be understood as the first term of a sum involving internal Elko productions of  the same type (a ``cascade'' of a ``fork''), what means that its contribution can be improved by the sum of those graphs, faced as a finite geometric series on $\lambda^2$; second, it should be stressed for completeness, that another factor resulting as an unexpected asymmetry on $\phi$ (for graphs involving four Elkos coupling) arises from the inclusion of the $\stackrel{\neg}{\eta}\stackrel{\neg}{\eta}$ and $\eta\eta$ type propagators, which are proportional to $N(p^\prime)$ and $M(p)$ matrices,  the ``twisted spin sums'':
\begin{eqnarray}\nonumber
M(p)&=&\left[\begin{array}{cccc}e^{-i\phi}p\,cos(\theta)&p\,sin(\theta)&0&-iE\\p\,sin(\theta)&-e^{i\phi}p\,cos(\theta)&iE&0\\0&-iE&-e^{-i\phi}p\,cos(\theta)&-p\,sin(\theta)\\-iE&0&-p\,sin(\theta)&e^{i\phi}p\,cos(\theta)\end{array}\right],\nonumber\\
N(p^\prime)&=&\left[\begin{array}{cccc}\sqrt{p'^2+m_{\varsigma}^2}&0&ip'sin(\theta^\prime)&-ie^{-i\phi^\prime}p'cos(\theta^\prime)\\0&\sqrt{p'^2+m_{\varsigma}^2}&-ie^{i\phi^\prime}p'cos(\theta^\prime)&-ip'sin(\theta^\prime)\\ip'sin(\theta^\prime)&-ie^{-i\phi^\prime}p'cos(\theta^\prime)&-\sqrt{p'^2+m_{\varsigma}^2}&0\\-ie^{i\phi^\prime}p'cos(\theta^\prime)&-ip'sin(\theta^\prime)&0&-\sqrt{p'^2+m_{\varsigma}^2}\end{array}\right].
\end{eqnarray}


\begin{acknowledgements}
We are grateful to professor D. V. Ahluwalia and professor A. Alves for valuable suggestions.
\end{acknowledgements}



\begin{thebibliography}{100}
\bibitem{elko}
D.~V.~Ahluwalia and D.~Grumiller,
JCAP {\bf 0507}, 012 (2005)
[arXiv:hep-th/0412080].

\bibitem{elko2}
D.~V.~Ahluwalia and D.~Grumiller,
Phys.\ Rev.\ D {\bf 72}, 067701 (2005)
[arXiv:hep-th/0410192].

\bibitem{COSMO} C. G. Boehmer, Annalen Phys. {\bf 16}, 325 (2007) [arXiv:gr-qc/0701087]; C. G. Boehmer, Annalen Phys. {\bf 16}, 38 (2007) [arXiv:gr-qc/0607088]; C. G. Boehmer and J. Burnett, Mod. Phys. Lett. {\bf A 25}, 101 (2010) [arXiv:0906.1351 [gr-qc]]; S. Shankaranarayanan, Int. J. Mod. Phys. {\bf D 18}, 2173 (2009) [arXiv:0905.2573 [astro-ph]]; C. G. Boehmer and J. Burnett, Phys. Rev. {\bf D 78}, 104001 (2008) [arXiv:0809.0469 [gr-qc]]; D. Gredat and S. Shankaranarayanan, JCAP {\bf 1001}, 008 (2010) [arXiv:0807.3336 [astro-ph]]; C. G. Boehmer, Phys. Rev. {\bf D 77}, 123535 (2008) [arXiv:0804.0616 [astro-ph]]; C. G. Boehmer and D. F. Mota, Phys. Lett. {\bf B663}, 168 (2008) [arXiv:0710.2003 [astro-ph]]; L. Fabbri, Mod. Phys. Lett. {\bf A 25 } [arXiv:0911.5304 [gr-qc]], 2483 (2010); L. Fabbri [arXiv:1008.0334 [gr-qc]].

\bibitem{MATE} L. Fabbri, Mod. Phys. Lett. {\bf A 25}, 151 (2010) [arXiv:0911.2622 [gr-qc]]; R. da Rocha and W. A. Rodrigues, Jr., Mod. Phys. Lett. {\bf A 21}, 65 (2006) [arXiv:math-ph/0506075]; R. da Rocha and J. M. Hoff da Silva, J. Math. Phys. {\bf 48}, 123517 (2007) [arXiv:0711.1103 [math-ph]]; R. da Rocha and J. M. Hoff da Silva, Adv. Appl. Clifford Alg., [arXiv:0811.2717 [math-ph]]; J. M. Hoff da Silva and R. da Rocha, Int. J. Mod. Phys. {\bf A 24}, 3227 (2009) [arXiv:0903.2815 [math-ph]].

\bibitem{WIG}
E.~P.~Wigner, Annals.\ Math.\ {\bf 40}, 149 (1939).

\bibitem{ALS}
D.~V.~Ahluwalia, C-Y~Lee, and D.~Schritt, and T. F. Watson, Phys.\ Lett.\ {\bf B 687}, 248 (2010) [arXiv:0804.1854  [hep-th]]; D.~V.~Ahluwalia, C-Y~Lee, and D.~Schritt, [arXiv:0911.2947 [hep-ph]].

\bibitem{MAG} K. Land and J. Magueijo, Mon. Not. Roy. Astron. Soc. {\bf 378}, 153 (2007) [arXiv:astro-ph/0611518]; P. K. Samal, R. Saha, P. Jain, and J. P. Ralston (2008) [arXiv:0811.1639 [astro-ph]]; M. Fommert and T. A. Ensslin (2009) [arXiv: 0908.0453 [astro-ph]].

\bibitem{VSR} A. G. Cohen and S. L. Glashow, Phys. Rev. Lett. {\bf 97}, 021601 (2006) [hep-ph/0601236].

\bibitem{SAL} D. V. Ahluwalia and S. P. Horvath, JHEP {\bf 11}, 078 (2010)   [arXiv:1008.0436 [hep-ph]]. 

\bibitem{Nath:2010zj}
P.~Nath {\em et~al.},
Nucl.\ Phys.\ B (Proc. Suppl.){\bf 200-202}, 185 (2010)
  [arXiv:1001.2693].

\bibitem{ATLAS} ATLAS inner detector: Technical design report. Vol. 1, CERN-LHCC-97-16.

\bibitem{Martin:1997ns}
S.~P.~Martin,
  [arXiv:hep-ph/9709356].

\bibitem{AguilarSaavedra:2005pw}
J.~A.~Aguilar-Saavedra{\em et~al.}
Eur.\ Phys.\ J{\bf C46}, 43 (2006),
  [arXiv:hep-ph/0511344].
\bibitem{CDFNote}CDF Note 10223
The CDF Collaboration,
Preliminary Results for the ICHEP 2010 Conference.


\bibitem{diagrammar}
  M.~J.~G.~Veltman,
{\it  Cambridge, UK: Univ. Pr. (1994) 284 p. (Cambridge lecture notes in physics, 4)}.
\bibitem{xloops}
  C.~Bauer and H.~S.~Do,
  Comput.\ Phys.\ Commun.\  {\bf 144}, 154 (2002)
  [arXiv:hep-ph/0102231].
\bibitem{cuba}
  T.~Hahn,
  Comput.\ Phys.\ Commun.\  {\bf 168}, 78 (2005)
  [arXiv:hep-ph/0404043].

\bibitem{:1994pu} CMS Physics Technical Design Report, Volume II: Physics Performance, J. Phys. G: Nucl. Part. Phys. 34, 995 (2007).

\bibitem{Cheung:2010af}
  K.~Cheung, J.~Song,
  Phys.\ Rev.\  {\bf D81}, 097703 (2010).
  [arXiv:1004.2783 [hep-ph]].

\bibitem{pdg}K. Nakamura et al. (Particle Data Group), J. Phys. G {\bf 37}, 075021 (2010). 


\end{thebibliography}
\end{document}